\documentclass[12pt]{article}
\begin{document}

\addtolength{\baselineskip}{0.5\baselineskip}

\title{\textbf{Quantum Chemistry at Finite Temperature}}
\author{Liqiang Wei\\
Chemical Physics Research Institute\\
Abington, MA 02351\footnote{wei@chemphysres.org}}

\maketitle

\begin{abstract}
\vspace{0.05in}
  In this article, we present emerging fields of quantum chemistry at finite temperature. We discuss its recent
 developments on both experimental and theoretical fronts. First, we describe several experimental investigations
 related to the temperature effects on the structures, electronic spectra, or bond rupture forces for molecules.
 These include the analysis of the temperature impact on the pathway shifts for the protein unfolding by atomic force
 microscopy ($AFM$), the temperature dependence of the absorption spectra of electrons in solvents, and the temperature
 influence over the intermolecular forces measured by the $AFM$.  On the theoretical side,
 we review advancements made by the author in the coming fields of quantum chemistry at finite temperature.
 Starting from the $\it{Bloch}$ equation, we have derived the sets of hierarchy equations for the reduced density
 operators in both canonical and grand canonical ensembles. They provide a law according to which the reduced density
 operators vary in temperature for the identical and interacting many-body systems. By taking the independent
 particle approximation, we have solved the equations in the case of a grand canonical ensemble, and obtained
 an energy eigenequation for the molecular orbitals at finite temperature. The explicit expression for the
 temperature-dependent $\it{Fock}$ operator is also given. They form a mathematical foundation for the examination of
 the molecular electronic structures and their interplay with finite temperature. Moreover, we clarify the physics
 concerning the temperature effects on the electronic structures or processes of the molecules, which is crucial for
 both theoretical understanding and computation. Finally, we summarize our discussion and point out some of the
 theoretical and computational issues in the fields of quantum chemistry at finite temperature for the future
 exploration.
\end{abstract}


\vspace{0.35in}
\section{Introduction}

 The history for quantum chemistry development is almost synchronous to that of quantum mechanics itself.
 It begins with $\it{Heitler}$ and $\it{London}$'s investigation of the electronic structure for $H_{2}$
 molecule shortly after the establishment of the wave mechanics for quantum particles~\cite{heitler}.
 There are two major types of molecular electronic theories: valence bond approach $\it{vs.}$ molecular orbital method
 with the latter being the popular one for the present study. It has gone through the stages from the evaluation
 of molecular integrals via a semiempirical way to the one by an $\it{ab\ initio}$ method. The correlation issue is
 always a bottleneck for the computational quantum chemistry and is under intensive examination for over fifty
 years~\cite{keinan}. For large molecular systems such as biomolecules and molecular materials, the development of the
 combined $\it{QM/MM}$ approach, the pseudopotential method and the linear scaling algorithm has significantly advanced
 our understanding of their structure and dynamics. There are about eight $\it{Nobel}$ prize laureates whose researches
 are related to the molecular electronic structure theory. This not only recognizes the most eminent scientists who have
 made the outstanding contributions to the fields of quantum chemistry, but more importantly, it indicates the essential
 roles the electronic structure theory has been playing in the theoretical chemistry as well as for the whole areas of
 molecular sciences~\cite{keinan}. Nowadays, quantum chemistry has been becoming a maturing
 science~\cite{wei1,wei2}.

 Nevertheless, the modern fields of quantum chemistry are only part of the stories for the molecular electronic
 structure theory. From the pedagogical points of view, the quantum mechanics based on which the traditional quantum
 chemistry is built is a special case of more general quantum statistical mechanics~\cite{wei3,wei4,lowdin2}. In
 reality, the experimental observations are made under the conditions with thermodynamic constraints. Henceforth,
 there is a need to extend the contemporary areas of quantum chemistry to the realm of, for instance, finite
 temperature~\cite{wei3,wei4,lowdin2}.

  Indeed, many experimental investigations in various fields and for different systems have already shown the
 temperature or pressure effects on their microscopic structures [8-31,50-58,64,72-88]. The polymeric molecule is one
 of the most interesting systems for this sort of studies [8-17]. The experimental measurement on the absorption
 spectra, photoluminescence ($\it{PL}$), and photoluminescence excitation ($\it{PLE}$), and spectral line narrowing
  ($\it{SLN}$) for the $\it{PPV}$ and its derivatives all show the same trend of the blue shift with an increasing
 temperature~\cite{hagler,friend,yu1}. This attributes to the temperature dependence of their very rich intrinsic
 structures such as the vibronic coupling~\cite{yu2,deleuze1,deleuze2}. The experimental inquiry of the
 temperature impacts on the biomolecules started in the late nineteenth century~\cite{rakow,waugh}. Most recently, it
 has been extended to the study of folding and unfolding of proteins or $\it{DNA}$s~\cite{gaub0,williams,law}. In
 addition to the observed patterns for the unfolding forces with respect to the extension or temperature, it has
 been proved that the temperature-induced unfolding is another way for the examination of mechanisms or pathways of
 protein folding or unfolding processes [20-24]. The newest related development is on the $\it{AFM}$ measurement made by
 $\it{Lo}$ et al. of the intermolecular forces for the biotin-avidin system in the temperature range from $286$ to
  $310 K$~\cite{simons}. It has shown that an increase of temperature will almost linearly decrease the strength of the
  bond rupture force for the individual biotin-avidin pair. The analysis of the temperature influences over the
  absorption spectra of the solvated electron began in the 1950's and it is still of current interest. A striking effect
  is that an increasing temperature will cause the positions of their maximal absorption red shift [72-86].

 In two papers recently published, we have deduced an energy eigenequation for the molecular orbitals~\cite{wei3,wei4}.
 It is the extension from the usual $\it{Hartree-Fock}$ equation at zero temperature to the one at any finite
 temperature~\cite{hartree,fock}. It opens an avenue for the study of the temperature impacts on the electronic
 structures as well as their interplay with the thermodynamic properties. In the third section, we will present this
 equation and give the details for its derivation. We will also expound the physics concerning the effects from
 temperature and classify them into two types. The one is at the single orbital level and the other is really an
 $N$-body effect. In the next section, we will show four major types of experiments related to the investigation of
 the temperature influences over the microscopic structures of molecular systems. In the final section, we will
 discuss and analyze our presentations, and point out both theoretical and computational issues for the future
 examination.

 \vspace{0.35in}
\section{Experimental Development}
 In this section, we focus our description of the experimental investigations related to the temperature
 effects on the bondings, structures, or electronic spectra of molecules. We choose four kinds of the most
 recent developments in these fields which are of chemical, biological, or material interests.

 \subsection{Temperature effects on geometric structure and UV-visible electronic spectra of polymers}

 The first important systems where the major issues related to the temperature influences over the geometric
 structures and electronic spectra are the polymeric molecules. Many experimental inquiries and some theoretical
 works already exist in the literature [8-17]. However, how temperature changes the microscopic structures of
 the polymers are still not completely understood and there are many unresolved problems in interpreting their
 electronic spectra. We list here a few very interesting experimental examinations for the purpose of demonstration.

 The poly($\it{p}$-phenylenevinylene)($\it{PPV}$) is one of the prototype polymeric systems for the study of their
 various mechanical, electronic, and optical properties. The impact from temperature on the absorption spectra,
 the photoluminescence ($\it{PL}$), and the photoluminescence excitation ($\it{PLE}$) of the $\it{PPV}$
 has also been investigated both experimentally and theoretically~\cite{hagler,friend,yu1}. In an experiment
 performed by $\it{Yu}$ et al., the absorption spectra are measured for the $\it{PPV}$ sample from the temperature
 $10$ to $330 K$. The details of the experiment are given in their paper~\cite{yu1}. The resulting spectra for the
 absorption at $T = 80$ and $300 K$ are shown in Figure 1 of that paper. We see that there
 is a pronounced change in the spectra when increasing the temperature. They also study the $\it{PL}$ and $\it{PLE}$
 spectra for the $\it{PPV}$. The measured $\it{PL}$ spectra at two temperatures: $77$ and $300 K$ are demonstrated
 in Figure 3, and the $\it{PLE}$ spectra at those temperatures are depicted in Figure 4 of the paper~\cite{yu1}.
 They both show the dramatic changes of the band blue shift when the temperature is increased. Similar studies have also
 been carried out before by the other groups~\cite{hagler,friend}. They observed the similar behaviors.

 Another interesting analysis is related to the temperature effects on the spectral line narrowing ($\it{SLN}$) of the
 poly(2-methoxy-5-($2^{'}$-ethylhexyloxy)-1,4-phenylenevinylene)($\it{MEH-PPV}$) spin-coated from either
 $\it{THF}$ or $\it{CB}$ solvents~\cite{sheridan}. In the experiment conducted by Sheridan et al., the $\it{SLN}$ is
 measured together with the absorption and $\it{PL}$ as shown in Fig. 1 of their paper. It is found that
the same trend of the $\it{SLN}$ blue shift is observed as that
for the absorption and $\it{PL}$ with an increasing temperature.
They attribute this to the same reason of the electronic structure
modification resulting from the variation of temperature.

 \subsection{Temperature effects on structure, dynamics, and
 folding/unfolding of biomolecules}
  Biomolecules are complex systems, featuring a large molecular size, a heterogeneity of atomic constitutes and
 a variety of conformations or configurations. Their energy landscapes thereby exhibit multiple substates and multiple
 energy barriers, and vary in size for the barrier heights~\cite{wolynes1,shakhnovich,karplus1,dill,thirumalai}.
 The temperature should have a strong influence over their structure and dynamics including the folding or
 unfolding [18-58]. This effect could be either from the fluctuation of thermal motions of the molecules or due to
 the redistribution of the electronic charge as we will discuss in the next section.

 The experimental observation of the temperature impact on the microscopic structure of biological systems dates back
 to the very early days. One focus, for example, is on the measurement of the elastic properties of the human red blood
 cell membrane as a function of temperature~\cite{rakow,waugh}. Another related investigation is about the influence
 over the thermal structural transition of the young or unfractionated red blood cells due to the involvement of the
 protein spectrin which might modify the spectrin-membrane interaction~\cite{minetti1,minetti2}. Most recently, the
 atomic force microscopy ($\it{AFM}$) has been used to detect the impact from the variation of temperature on the
 spectrin protein unfolding force as well as on the bond rupture force for the biotin-avidin
 system~\cite{gaub0,williams,law,simons}.

 The $\it{AFM}$ is a surface imaging technique with an atomic-scale resolution capable of measuring $\it{any}$ types of
 the forces as small as $10^{-18}\ N$. It combines the principles of the scanning tunnelling microscopes ($STM$) and
 the stylus profilometer, and therefore can probe the surfaces of both conducting and nonconducting
 samples~\cite{binnig1,binnig2}. The imaging on soft materials such as biomolecules with the $\it{AFM}$ has been
 accomplished beginning in the 1980's~\cite{hansma1,hansma2,gaub1}. Recently, it has been applied to measure the
 adhesive forces and energies between the biotin and avidin pair as we will show in the next
 subsection~\cite{moy1,moy2,moy3,beebe1}. Unlike other experimental techniques, the $\it{AFM}$ features a high precision
 and sensitivity to probe the surface with a molecular resolution, and can be done in physiological environments.

 In an $\it{AFM}$ investigation of the mechanical unfolding of titin protein, for example, the restoring forces all
 show a sawtooth like pattern with a definite periodicity. It reveals much information about the mechanism of the
 unfolding processes~\cite{rief,fernandez1}. The observed pattern, in addition to a fit of a worm-like chain model,
 has also been verified by the steered molecular dynamics or Monte carlo simulations~\cite{schulten1,discher2}.
 Similar study has been extended to other systems~\cite{fernandez2,lenne,discher3}.

 The same kind of experiments has also been performed by varying temperature. In the experiment carried out by Spider
  and Discher et al.~\cite{law}, the spectrin protein is chosen for the $\it{AFM}$ study at different temperatures.
 Thousands of tip-to-surface contacts are conducted for a given temperature because of the statistical nature of the
 $\it{AFM}$ measurement. The observed curve for the relation between the unfolding force and extension shows the
 similar sawtooth pattern for all temperatures. In addition, the tandem repeat unfolding events are more favored at
 lower temperature as demonstrated in the unfolding length histograms. Most striking is that the unfolding forces
 show a dramatically nonlinear decreasing relation as the temperature $T$ approaches the transition temperature $T_{m}$.
 This is shown in Figure $3B$ of the paper~\cite{law}.

 Similar behaviors regarding the force-temperature dependence have also been observed via either $\it{AFM}$ or
 optical tweezers for the forced overstretching transition for the individual double-stranded $\it{DNA}$
 molecules~\cite{gaub0,williams,bloomfield1,bloomfield2}.

 Some other interesting experiments which illustrate the effects from temperature on the microscopic structures of
 biomolecules have also been performed even though the detailed physical origins of the impacts (from either the
 electrons or the molecules) have not been specified~\cite{eriksson,huang,wang1,wang2,mayer,dyer}. In a circular
 dichroism ($\it{CD}$) spectra and high resolution $\it{NMR}$ study, for instance, it shows that the secondary
 structure of the Alzheimer $\beta$ (12-28) peptide is temperature-dependent with an extended left-handed
 $3_{1}$ helix interconverting with a flexible random coil conformation~\cite{eriksson}. Another example
 is related to the analysis of the temperature-dependent interaction of the protein $\it{Ssh10b}$ with a $\it{DNA}$ which
 influences the $\it{DNA}$ topology~\cite{huang,wang1,wang2}. The analysis from the heteronuclear $\it{NMR}$ and
 site-directed mutagenesis indicates that the $\it{Ssh10b}$ exists as a dimer: $T$ form and $C$ form. Their ratio is
 determined by the $Leu^{61}-Pro^{62}$ peptide bond of the $\it{Ssh10b}$ which is sensitive to temperature.

 \subsection{Temperature effects on intermolecular forces}
 The analysis of the general issues related to the temperature effects on the microscopic structures has been most
 recently extended to the realm of intermolecular forces. Since the usual intermolecular forces
  such as hydrogen bonds, $\it{van\ der\ Waals}$ forces, ionic bonds, and hydrophobic interactions are weak and typically
  of the order of $\it{0.1\ eV}$ or $\it{4.0\ kT}$ at the physiological temperature,
  the variation of temperature will thereby have a very strong influence over the strength of these forces.

 The first experimental investigation on the temperature-dependent intermolecular forces is for the
 biotin-avidin system and by an $AFM$ measurement~\cite{simons}. The biotin-avidin complex is a prototypical receptor
 and ligand system with the biotin binding strongly up to four avidin
 proteins~\cite{green,darst,sussman,bolognesi1,bolognesi2}. They have an extremely high binding affinity, and therefore
 serves as a model system for various experimental examinations.
 In the experiment carried out in $\it{Beebe}$'s group~\cite{simons}, the receptor avidin
 is attached to the  $\it{AFM}$ tip and linked to the agarose bead functionalized with the biotin.
  The temperature of the entire $\it{AFM}$ apparatus is varied at a range from
  $286$ to $310 K$. In addition, the loading rate is kept very slow so that the thermal equilibrium for the
  biotin-avidin pairs is assumed. The forces expected to be determined is the rupture force $F_{i}$ between
  the individual biotin-avidin pair which is defined as the maximum restoring force~\cite{simons}. In an actual $AFM$
  experiment, however, the total adhesive force between the tip and substrate is measured. It is a
  sum of the finite number $n$ of the interactions between each biotin and avidin pair. To extract the individual and
  average bond rupture, a statistical method has been developed in $\it{Beebe}$'s
 group~\cite{beebe1,beebe2,beebe3,beebe4,bai1}. They assume a $\it{Poisson}$ distribution for the number $n$ of the
 discrete rupture forces or linkages from multiple measurements, and have obtained the single force $F_{i}$ at
  different temperatures. The result is shown in Figure $3$ of the paper~\cite{simons}. We see that the individual
  rupture force $F_{i}$ for the biotin-avidin pair is decreased by about five-fold in strength when the temperature is
  increased from $286$ to $310 K$.

  To interpret the observed temperature impact on the biotin-avidin forces, $\it{Peebe}$'s group has performed a
 thermodynamic analysis~\cite{simons}. Based on the simple models and arguments~\cite{tavan,schulten}, they have
 come out an equation that connects the square of the single bond-rupture force $F_{i}$ to the absolute temperature
 $T$ as follows,
 \begin{equation}
  F^{2}_{i} = 2\Delta E^{\ddagger} k_{bond} - 2k_{B} T k_{bond}
  \ln\left(\frac{\tau_{R}}{\tau_{D}}\right)
\end{equation}
 where the $k_{bond}$ is the force constant of the individual biotin-avidin pair, and the time $\tau_{R}$ is the
 characteristic time needed to break $n$ pairs of those forces. The $E^{\ddagger}$ is the energy required to remove
 the biotin from avidin's strongest binding site and the corresponding time is $\tau_{D}$.  More details on this
 analysis can be found in the paper~\cite{simons}. The relation between the square of the force $F_{i}$ and the temperature
 $T$ is also plotted as Figure $5$ in that paper. Therefore, from the relation (1) and this figure, the information
  about the stiffness of the ligand and receptor bond and the critical binding energy, etc. can be obtained.
 Obviously, what we need at present is a microscopic theory which can account for all these relations and properties.

\subsection{Temperature effects on absorption spectra of electrons in solvents}

 The structure and dynamics of the solute in solvents is one of the most important fields in chemistry since most of the
 chemical reactions occur in solution phases. In the meantime, it is also one of the most challenging areas in
 theoretical chemistry with many unsettled issues. The variation of
 temperature in the measurement of absorption spectra of solvated electron in various solvents has proved to be
 a useful means for the understanding of the solvation processes [71-85].

 There are several experimental techniques available for this type of inquires with the pulse radiolysis being the one
 most commonly used. There are also several research groups conducting the similar experimental investigations and
 obtaining the consistent results relating to the temperature effects on the optical absorption spectra of the solvated
 electron in solvents. In a recent experiment carried out in $\it{Katsumura}$'s group, for example, the pulse radiolysis
 technique is employed to study the optical absorption spectra of the solvated electron in the ethylene glycol at
 different temperatures from $290$ to $598 K$ at a fixed pressure of $100$ atm. In addition to the faster
 decay of absorptions, it is found that, their maximal positions shift to the red with the increasing temperature as
 shown in Figures 1 to 3 of the paper~\cite{katsumura4}.
 This is in contrast to the situation for the electronic spectra of the polymers. They also point out the need to
 quantify the change of the density in the experiment in order to really understand the observed results.

 The same type of experiments has been extended to the examination of the optical absorption spectra for $Ag^{0}$ and
  $Ag_{2}^{+}$ in water by varying temperature, and similar results have been obtained~\cite{katsumura3}.

 \vspace{0.35in}
\section{Theoretical Development}

 Having presented four different types of experiments above, we can observe that the investigation of the temperature
 effects on the microscopic structures of molecules is a very interesting and sophisticated field. More need to be probed
 and understood. Even though the experimental analysis has been for a long time, very limited number of the related
 theoretical works is available, especially at the first-principle level.
 In other words, the quantum chemistry at finite temperature is not a well-established field~\cite{wei3,wei4,lowdin2}.

 It is true that the influence of temperature on the microscopic structures is a complicated phenomenon. There exists
 different functioning mechanisms. One consideration is that the variation of temperature, according to the Fermi-Dirac
 statistics, will change the thermal probability distribution of single-particle states for a free electron gas. It is
 expected that similar situation should occur for an interacting electron system, and therefore its microscopic
 structure will be correspondingly altered. Another consideration is that, for molecules or solids, the thermal
 excitation will cause the change of the time scales for the molecular motions. This will most likely bring about the
 transitions of the electronic states, and therefore lead to the breakdown of the Born-Oppenheimer approximation.
 The electron-phonon interaction is a fundamental topic in solid state physics and its temperature dependence is
 well-known. As a result, the variation of temperature will change the strength of the coupling between the electronic
 and molecular motions. Nevertheless, we tackle the issues pertaining to the temperature effects in a simpler way. We
 treat only an identical and interacting $\it{fermion}$ system. Or we neglect the coupling of the electronic motion
 with those of the nucleus in the molecules or solids. We expect that some sort of the general conclusions will come
 out from this analysis. As a matter of fact, this is the approach commonly admitted in a non-adiabatic molecular
 dynamics study, in which purely solving the eigenequation for the electrons will provide the reference states for the
 examination of the coupling between the electronic and nuclear motions of the molecules.

 In the following, we will present self-consistent eigenequations within the framework of the density operators in
 equilibrium statistical mechanics which decides the molecular orbitals at a given temperature.

 \subsection{Hierarchy Bloch equations for reduced density
 operators in canonical ensemble}
 We consider an identical and interacting $N$-particle system. In a canonical ensemble, its
 $N$th-order density operator takes the form
 \begin{equation}
   D^{N} = \exp(-\beta H_{N}),
\end{equation}
 and satisfies the $\it{Bloch}$ equation~\cite{bloch,kirkwood}
 \begin{equation}
-\frac{\partial}{\partial\beta} D^{N} = H_{N} D^{N},
\end{equation}
 where
 \begin{equation}
  H_{N} = \sum_{i=1}^{N} h(i) + \sum_{i<j}^{N} g(i,j),
\end{equation}
 is the $\it{Hamiltonian}$ for the $N$ particle system composed of one-particle operator $h$ and two-body
 operator $g$. The $\beta$ is the inverse of the product of $\it{Boltzmann}$ constant $k_{B}$ and absolute
  temperature $T$.

 Since the $\it{Hamiltonian}$ (4) can be written as a reduced $\it{two}$-body operator form,
 the second-order reduced density operator suffices to describe its $N\ (\ge 2)$ particle quantum states. A $p$th-order
 reduced density operator is generally defined
 by~\cite{kummer,harriman1}
 \begin{equation}
D^{p} = L^{p}_{N} (D^{N}),
\end{equation}
 where $L^{p}_{N}$ is the contraction operator acting on an $N$th-order tensor
 in the $N$-particle $\it{Hilbert}$ space $V^{N}$. The trace of the $D^{p}$ gives the partition
 function,
\begin{equation}
 Tr (D^{p}) = Z (\beta, V, N).
\end{equation}
  Rewrite the $\it{Hamiltonian}$ in a form
 \begin{equation}
H_{N}=H^{p}_{1}+\sum_{j=p+1}^{N}h(i)+\sum_{i=1}^{p}\sum_{j=p+1}^{N}g(i,j)
+\sum_{i<j\\(i\ge p+1)}^{N} g(i,j),
\end{equation}
where
\begin{equation}
  H_{1}^{p} = \sum_{i=1}^{p} h(i) + \sum_{i<j}^{p} g(i,j),
\end{equation}
 and apply the contraction operator $L^{p}_{N}$ on both sides of the Eq. (3), we develop
 an equation that the $p$th-order density operator
 satisfies~\cite{wei3}
 \begin{eqnarray}
\nonumber
-\frac{\partial}{\partial\beta}D^{p}&=&H^{p}_{1}D^{p}+(N-p)L^{p}_{p+1}
\left[h(p+1)D^{p+1}\right]+(N-p)L_{p+1}^{p}\left[\sum_{i=1}^{p}g(i,p+1)D^{p+1}\right]+\\
&&+\left(\begin{array}{c} N-p\\2\end{array}\right)
L_{p+2}^{p}\left[g(p+1,p+2)D^{p+2}\right].
\end{eqnarray}
 It provides a law according to which the reduced density operators vary
 in terms of the change of temperature.

 \subsection{Hierarchy Bloch equations for reduced density
 operators in grand canonical ensemble}
 The above scheme for deducing the equations for the reduced operators can be readily extended to
  the case of a grand canonical ensemble. It is a more general one with a
 fluctuating particle number $N$. In this ensemble, the density operator is defined in the entire
  $\it{Fock}$ space
 \[ F = \sum_{N=0}^{\infty} \oplus V^{N},   \]
  and is written as the direct sum of the density operators $D_{G}(N)$
 associated with the $N$-particle $\it{Hilbert}$ space $V^{N}$,
\begin{equation}
   D_{G} = \sum_{N=0}^{\infty} \oplus D_{G}(N),
\end{equation}
 where
 \begin{eqnarray}
\nonumber D_{G}(N)&=& \exp[-\beta(H-\mu N)],\\
  &=& \exp(-\beta \bar{H}),
\end{eqnarray}
and
\begin{equation}
  \bar{H} = H - \mu N,
\end{equation}
is called the grand $\it{Hamiltonian}$ on $V^{N}$. The form of the
$\it{Hamiltonian}$ $H$ has been given by Eq. (4) and the $\mu$ is
the chemical potential. The corresponding $p$th-order reduced
density operator is therefore defined as
\begin{equation}
 D^{p}_{G} =\sum_{N=p}^{\infty}\oplus
 \left(\begin{array}{c}N\\p\end{array}\right)L_{N}^{p}[D_{G}(N)]
\end{equation}
with the trace given by
\begin{equation}
  Tr(D_{G}^{p}) =
  \left<\left(\begin{array}{c}N\\p\end{array}\right)\right>
  D^{0}_{G},
\end{equation}
 and
\begin{equation}
  D_{G}^{0} = \Xi(\beta, \mu, V).
\end{equation}
 The $\Xi(\beta, \mu, V)$ is the grand partition function.

 In a similar manner, we can also derive the hierarchy equations that
 the reduced density operators in the grand canonical ensemble
 obey~\cite{wei4}
\begin{eqnarray}
\nonumber
-\frac{\partial}{\partial\beta}D^{p}&=&\bar{H}^{p}_{1}D^{p}+(p+1)L^{p}_{p+1}
\left[\bar{h}(p+1)D^{p+1}\right]+(p+1)L_{p+1}^{p}\left[\sum_{i=1}^{p}g(i,p+1)D^{p+1}\right]+\\
&&+\left(\begin{array}{c} p+2\\2\end{array}\right)
L_{p+2}^{p}\left[g(p+1,p+2)D^{p+2}\right],
\end{eqnarray}
where
\begin{equation}
  \bar{H}_{1}^{p} = \sum_{i=1}^{p}\bar{h}(i) + \sum_{i<j}^{p}
  g(i,j),
\end{equation}
and
\begin{equation}
 \bar{h}(i) = h(i)-\mu.
\end{equation}
It gives us a law with which the reduced density operators in the
grand canonical ensemble vary in temperature.

 \subsection{Orbital approximation and Hartree-Fock equation at
  finite temperature}
  The Eqs. (9) and (16) define a set of $\it{hierarchy}$ equations that establish the relation among the reduced density
 operators $D^{p}$, $D^{p+1}$, and $D^{p+2}$. They can be solved either in an exact scheme or by an approximate
 method. The previous investigation of $N$ electrons with an independent particle approximation to the
 $\it{Schr}$$\ddot{o}$$\it{dinger}$ equation for their $\emph{pure}$ states has lead to the $\it{Hartree-Fock}$ equation
 for the molecular orbitals [95-101]. We thereby expect that the same approximate scheme to the reduced $\it{Bloch}$
 equations (9) or (16), which hold for more general mixed states, will yield more generic eigenequations than the usual
  $\it{Hartree-Fock}$ equation for the molecular orbitals.

 We consider the case of a grand canonical ensemble. For $p = 1$, Eq. (16) reads
\begin{equation}
-\frac{\partial}{\partial\beta}
D^{1}=\bar{H}_{1}D^{1}+\frac{Tr(\bar{h}D^{1})}{D^{0}}D^{1}-\frac{1}{D^{0}}D^{1}\bar{h}
D^{1}+2L^{1}_{2}\left[g(1,2)D^{2}\right]+3L^{1}_{3}\left[g(2,3)D^{3}\right].
\end{equation}
  Under the orbital approximation, above second-order and third-order reduced density
  operators for the electrons can be written as
  \begin{equation}
\nonumber D^{3} = D^{1}\wedge D^{1}\wedge D^{1}/(D^{0})^{2}
 \end{equation}
  and
 \begin{equation}
 D^{2} = D^{1}\wedge D^{1}/D^{0}.
\end{equation}
These are the special situations for the statement that a
$p$th-order reduced density matrix can be expressed as a $p$-fold
 $\it{Grassmann}$ product of its first-order reduced density matrices.
 With this approximation, the last two terms of Eq. (19) can be evaluated in a straightforward way as
follows
\begin{equation}
 2L^{1}_{2}\left[g(1,2)D^{2}\right] = (J-K)D^{1},
\end{equation}
and
\begin{equation}
 3L^{1}_{3}\left[g(2,3)D^{3}\right]
 =\frac{Tr(gD^{2})}{D^{0}}-\frac{1}{D^{0}}D^{1}(J-K)D^{1},
\end{equation}
where
\begin{equation}
J = Tr_{2}\left[g\cdot D^{1}(2;2)\right]/D^{0},
\end{equation}
and
\begin{equation}
 K = Tr_{2}\left[g\cdot (2,3)\cdot D^{1}(2;2)\right]/D^{0},
\end{equation}
are called the $\it{Coulomb}$ and exchange operators,
respectively. With $(2,3)$ being the exchange between the particle
$2$ and $3$, the action of the $K$ on the reduced density operator
is
\begin{eqnarray}
\nonumber K\cdot D^{1}(3;3)&=& Tr_{2}\left[g\cdot (2,3)\cdot
D^{1}(2;2)\right]
/D^{0}\cdot D^{1}(3;3) \\
&=& Tr_{2}\left[g\cdot D^{1}(3;2)\cdot D^{1}(2;3)\right]/D^{0}.
\end{eqnarray}
Substitution of Eqs. (22) and (23) into Eq. (19) results in the
$\it{Bloch}$ equation for the first-order reduced density matrix
of the $N$ interacting electrons under the orbital approximation,
\begin{equation}
-\frac{\partial}{\partial\beta}D^{1}=(F-\mu)D^{1}+
 \left(\frac{Tr\ \bar{h}D^{1}}{D^{0}}
 +\frac{Tr\ g D^{2}}{D^{0}}\right)D^{1}-\frac{1}{D^{0}}D^{1}(F-\mu)D^{1},
\end{equation}
where
\begin{equation}
    F = h+J-K,
\end{equation}
is called the $\it{Fock}$ operator at finite temperature. Redefine
the normalized first-order reduced density operator
\begin{equation}
 \rho^{1} = D^{1}/D^{0},
\end{equation}
we can simplify above equation into
\begin{equation}
-\frac{\partial}{\partial\beta} \rho^{1}
=(F-\mu)\rho^{1}-\rho^{1}(F-\mu)\rho^{1}.
\end{equation}

Furthermore, from Eq. (30) and its conjugate, we get
\begin{equation}
  F\rho^{1}-\rho^{1}F = 0,
\end{equation}
 which indicates that the $\it{Fock}$ operator $F$ and the first-order reduced density matrix $\rho^{1}$ commute.
 They are also $\it{Hermitian}$, and therefore have common eigenvectors $\{|\phi_{i}\rangle\}$. These vectors are
 determined by the following eigen equation for the $\it{Fock}$ operator,
\begin{equation}
  F|\phi_{i}\rangle = \epsilon_{i}|\phi_{i}\rangle.
\end{equation}
 It is the eigenequation for the molecular orbitals at finite
 temperature.

The first-order reduced density operator is correspondingly
expressed as
\begin{equation}
  \rho^{1} =
  \sum_{i}\omega(\beta,\mu,\epsilon_{i})|\phi_{i}\rangle\langle\phi_{i}|,
\end{equation}
where $\omega(\beta,\mu,\epsilon_{i})$ is the thermal probability
that the orbital is found to be in the state
$\{|\phi_{i}\rangle\}$ at finite temperature $T$. Substituting Eq.
(33) into Eq. (30), we can obtain the equation this thermal
probability $\omega(\beta,\mu,\epsilon_{i})$ satisfies,
\begin{equation}
-\frac{\partial}{\partial\beta} \omega(\beta, \mu, \epsilon_{i})=
(\epsilon_{i}-\mu)\omega(\beta,\mu,\epsilon_{i})-(\epsilon_{i}-\mu)\omega^{2}
(\beta,\mu,\epsilon_{i}).
\end{equation}
Its solution has the same usual form of the $\it{Fermi-Dirac}$
statistics for the free electron gas as follows,
\begin{equation}
\omega(\beta,\mu,\epsilon_{i})=\frac{1}{1+e^{\beta(\epsilon_{i}-\mu)}},
\end{equation}
with the energy levels $\{\epsilon_{i}\}$ determined by Eq. (32).

\vspace{0.35in}
\section{Discussion, Summary and Outlook}
In this paper, we have presented both experimental and theoretical
developments related to the temperature impacts on the microscopic
structures and processes for the molecules.

In the theoretical part, we have depicted the sets of hierarchy
$\it{Bloch}$ equations for the reduced statistical density
operators in both canonical and grand canonical ensembles for the
identical fermion system with a two-body interaction. We have
solved the equations in the latter case under a single-orbital
approximation and gained an energy eigenequation for the
single-particle states. It is the extension of usual
$\it{Hartree-Fock}$ equation at absolute zero temperature to the
one at any finite temperature. The average occupation number
formula for each single-particle state is also obtained, which has
the same analytical form as that for the free electron gas with
the single-particle state energy determined by the
$\it{Hartree-Fock}$ equation at finite temperature (32).

From Eqs. (24), (25) and (28), we see that the $\it{Coulomb}$
operator $J$, the exchange operator $K$, and therefore the
$\it{Fock}$ operator $F$ are both coherent and incoherent
superpositions of the single-particle states. They are all
temperature-dependent through an incoherent superposition factor,
the $\it{Fermi-Dirac}$ distribution,
$\omega(\beta,\mu,\epsilon_{i})$. Therefore, the mean force or the
force field, and the corresponding microscopic structures are
temperature-dependent.

 We have expounded the physics relating to the temperature effects on the electronic structures or processes
 for the molecules. These effects can be either at the single-electron level or of the $N$-body excitation. This is
 very critical for our understanding and computation of the temperature influences over the molecular structures.
 From this analysis, for example, we can conclude that the temperature should
 have a stronger effect on the transition states for the molecules. Accordingly, the change of the chemical reactivity
 might come from the alternation of the electronic states due to the variation of temperature. We expect that this sort
  of the changes involving configuration mixing will be very common for the systems we are discussing here.
  Therefore, it will be a very significant work to develop or test the corresponding multireference theory for the
  molecular orbitals or electronic structures at finite temperature~\cite{keinan,wahl,gilbert,hirao,roos}. More general
  or standard theoretical explorations are expected~\cite{wei1}.

 On the experimental side, we have exposed four major fields of investigation of chemical, biomolecular, or material
 importance, which demonstrate the temperature impacts on their structures, spectra, or bond rupture forces.

 The complete determination of the geometric structures and electronic spectra of the polymeric molecules is a very
 difficult task. As has been stated in papers~\cite{deleuze1,deleuze2}, there are many different elements
 contributing to the change of the spectra. At present, we focus on the examination of the effect from temperature.
 We have showed that it can alter both the shapes and positions of the absorption and other spectra for
 the $\it{PPV}$ and its derivatives. As has been analyzed, the increase of temperature will bring about the excitation
 of the vibrational, rotational and liberal motions, which might also lead to the electronic transition. The
 $\it{Huang-Rhys}$ parameter has been introduced to describe the strength of the coupling between the electronic
 ground- and excited-state geometries. Furthermore, it has been observed that this factor is an increasing function
 of temperature~\cite{hagler,friend,yu1}. Obviously, a more detailed analysis of the electronic structure,
 excitation and spectroscopic signature at the first-principle level, which includes the temperature-dependent
 force field, is anticipated.

 Temperature has proved to be a big player in both experimental and theoretical study of the structure and
 dynamics of biomolecules including their folding or unfolding. At a first glance, the energy gap between the
 $\it{HOMO}$ and $\it{LUMO}$ for the biomolecules should be small or comparable to
 the $\it{Boltzmann}$ thermal energy $k_{B}T$ because of their very large molecular size. Therefore a change of
 temperature should have a strong influence over their electronic states, and consequently, the energy landscape and
 the related dynamics including the folding and unfolding. The experimental investigation with the $\it{AFM}$ and
 other techniques of the temperature effect on the shift of their unfolding pathways might have verified this type of
 thermal deformation of the potential energy landscape~\cite{gaub0,williams,law}. This is in contrast to the tilt and
 deformation of the energy landscape including its transition states for the biomolecules resulting from the applied
 mechanical forces~\cite{evans1}.

 Unfolding proteins by temperature is not just one of the classical experimental techniques for the analysis of the
 structure, dynamics and energetics of the biomolecules. It has also been utilized, for example, in the molecular
 dynamic simulation to study the structure of the transition states of $\it{CI2}$ in water at two different temperatures:
 $298\ K$ and $498\ K$~\cite{daggett}. The later high temperature is required in order to destabilize the native state
 for monitoring the unfolding as done in the real experiments. In another recent molecular dynamics
 simulation~\cite{karplus2}, $\it{Karplus}$'s group has compared the temperature-induced unfolding with the
 force-stretching unfolding for two $\beta$-sandwich proteins and two $\alpha$-helical proteins. They have found that
 there are significant differences in the unfolding pathways from two approaches. Nevertheless, in order to get more
 reliable results, the temperature-dependent force fields need to be developed and included in the molecular dynamics
 simulations. This is also the case in the theoretical investigation of protein folding since an accurate simulation of
 protein folding pathways requires better stochastic or temperature-dependent potentials which have become the bottleneck in structure
 prediction~\cite{wolynes1,shakhnovich,karplus1,dill,thirumalai}. From structural points of view, the variation of
 temperature leads to the change of the mean force or the energy landscape, and therefore provides a vast variety of
 possibilities, for instance, in the protein design and engineering.

 The intermolecular forces are ubiquitous in nature. They are extremely important for the biological systems and for
 the existence of life. The intermolecular forces have the specificity which is responsible for the molecular
 recognition between the receptor and ligand, the antibody and antigen, and complementary
 strands of $\it{DNA}$, and therefore for the regulation of complex organization of life~\cite{frauenfelder}.
 For these reasons, the experiment carried out in $\it{Beebe}$'s group has an immediate significance. It has
 demonstrated that temperature can be an important factor for changing the specificity of the intermolecular forces
 and therefore the function of life~\cite{simons}. Nevertheless, how the charge redistribution occurs due to the
 variation of temperature has not been interpreted, and a microscopic theory for quantifying the temperature
 influence over the intermolecular forces is still lacking. Since the delicate study of the intermolecular forces
 provides the insight into complex mechanisms of ligand-receptor binding and unbinding processes or pathways, a
 paramount future research is to establish the links between the intermolecular forces and the temperature within
 quantum many-body theory.

 The theoretical exploration of the temperature effects on the optical absorption spectra of solvated electrons is still
 in very early stage and few published works are available~\cite{jortner,brodsky,berne,nicolas}. One of the earliest
 inquires by $\it{Jortner}$ used a cavity model to simulate the solvated electron where the electron is confined to
 the cavity surrounded by the dielectric continuum solvent~\cite{jortner}. However, his investigation is not of fully
 microscopic in nature since he assumed a temperature dependence of phenomenological dielectric constants
 which were obtained from the available experimental data. In addition, the model used is too simplified
 and, for instance, it neglects the intrinsic structure of solvent molecules. There are a few recent examinations on
 the temperature influences over the absorption spectra of the solvated electrons. They all cannot catch the
 full features of the experimental observations. One reason is that the physical nature for the process is not totally
 understood which might leads to incorrect models employed for the simulation. The other is to utilize the crude models
 which might have omitted some important physical effects. For example, in an analysis by $\it{Brodsky}$ and
 $\it{Tsarevsky}$~\cite{brodsky}, they have concluded a temperature-dependence relation for the
 spectra which is, however, in contradiction with the experimental findings at high temperature. The quantum
 path-integral molecular dynamics simulation cannot produce those temperature-dependence relations observed in the
 experiments~\cite{berne}. In a recent quantum-classical molecular-dynamics study by $\it{Nicolas\ et\ al}$, even though
 the temperature-dependent features of optical absorption spectra for the solvated electron in water have been
 recovered~\cite{nicolas}, however, they claim that the red shifts of absorption spectra with the increasing temperature
 observed in both experiments and calculations are due to the density effect instead of temperature. This might cast the
 doubt of usefulness of our present theoretical work in this area. However, after examining their work, we observe that
 they actually have $\it{not}$ included any temperature effects on the electron in their theoretical model. These
 effects might be either from the $\it{Fermi-Dirac}$ distribution for individual electrons or due to the electronic
 excitation caused by the thermal excitation of the solvent, as we have discussed previously and in paper~\cite{wei4}.
 Obviously, much finer theoretical works or more experimental investigations in this area are expected to resolve this
 dispute.

 In addition to the systems discussed above, there are many other types which show the temperature impacts on their
 microscopic structures. Either theoretical or experimental works have been done or are in progress. Examples include
 the study of the temperature dependence of the $\it{Coulumb}$ gap and the density of states for the $\it{Coulomb}$
 glass, the experimental investigation of the temperature effects on the band-edge transition of $ZnCdBeSe$, and the
 theoretical description of the influence from temperature on the polaron band narrowing in the oligo-acene
 crystals~\cite{schreiber,hsieh,hannewald}.

  To sum up, the quantum chemistry at finite temperature is a new and exciting field. With the combination of the
  techniques from modern quantum chemistry with those developed in statistical or solid state physics, it will provide
  us with a myriad number of opportunities for the exploration~\cite{wei1,wei4,hirao,roos,wei6}\footnote{The theoretical
  development demonstrated in Section 3 of this paper was based on the thesis of the author in partial
  fulfillment of the requirements for his graduate degree at $Jilin$ University in June 1989. The original version of
  this paper as shown in physics/0412180 was finished in June 2004 when the author was a postdoc fellow in ITAMP at
  Harvard University.}




\end{document}